\documentclass{article}
\usepackage{spconf,amsmath,graphicx}
\usepackage{amssymb}
\usepackage{acronym}
\usepackage{url}
\usepackage{xcolor}
\usepackage{flushend} 
\usepackage[citecolor=black,
            colorlinks=true,
            linkcolor=black,
            urlcolor  = black,
            pdftitle={Multi-CMGAN+/+: Leveraging Speech Quality Metric Prediction for Training on Real Data For Speech Enhancement},
            pdfauthor={George Close, William Ravenscroft, Thomas Hain, and Stefan Goetze},
            pdfkeywords={speech enhancement, model generalisation, generative adversarial networks, conformer, metric prediction},
            pdfsubject={speech enhancement, model generalisation, generative adversarial networks, conformer, metric prediction}]{hyperref}
\usepackage[english]{babel}
\addto\extrasenglish{}
\addto\extrasenglish{}
\addto\extrasenglish{}
\addto\extrasenglish{}
\addto\extrasenglish{}
\addto\extrasenglish{}

\acrodef{GAN}  {generative adversarial network}
\acrodef{UDASE} {unsupervised domain adaptation speech enhancement}
\acrodef{CMGAN} {conformer-based metric \ac{GAN}}
\acrodef{SSSR}{Self Supervised Speech Representation}
\acrodef{BLSTM}{bi-di\-rec\-tional long short-term memory}
\acrodef{STFT}{short-time Fourier transform}
\acrodef{ISTFT}{inverse short-time Fourier transform}
\acrodef{PESQ}{Perceptual Evaluation of Speech Quality}
\acrodef{DNSMOS}{Deep Noise Suppression Mean Opinion Score}
\acrodef{MOS}{Mean Opinion Score}
\acrodef{SI-SDR}{scale invariant signal-distortion ratio}

\title{Multi-CMGAN+/+: Leveraging Multi-Objective Speech Quality Metric Prediction for Speech Enhancement }
\name{George Close, William Ravenscroft, Thomas Hain, and Stefan Goetze\thanks{This work was supported by the Centre for Doctoral Training in Speech and Language Technologies (SLT) and their Applications funded by UK Research and Innovation [grant number EP/S023062/1]. This work was also funded in part by TOSHIBA Cambridge Research Laboratory and 3M Health Information Systems Inc.}}
\address{Speech and Hearing Group, The University of Sheffield, Sheffield, UK}
\begin{document}
%
\maketitle
\begin{abstract}
  Neural network based approaches to speech enhancement have shown to be particularly powerful, being able to leverage a data-driven approach to result in a significant performance gain versus other approaches. Such approaches are reliant on artificially created labelled training data such that the neural model can be trained using intrusive loss functions which compare the output of the model with clean reference speech. Performance of such systems when enhancing real-world audio often suffers relative to their performance on simulated test data. In this work, a non-intrusive multi-metric prediction approach is introduced, wherein a model trained on artificial labelled data  using inference of an adversarially trained metric prediction neural network.  The proposed approach shows improved performance versus state-of-the-art systems on the recent CHiME-7 challenge \ac{UDASE} task evaluation sets. 
\end{abstract}

\section{Introduction}
For training of supervised neural-network based speech enhancement systems, there is often a mismatch between the synthetic data used to train the system and real-world recordings. This can lead to poor performance of such systems \emph{in the wild} even if intrusive evaluation metrics on synthetic data are high. A compounding factor in this problem is that metrics which are designed to measure speech quality do not always correlate strongly with actual human assessment of speech audio quality in many scenarios \cite{RGHKK08,GARHK13}, and often require access to clean reference/label audio which may not be readily available for real-life recordings.\\
Recently, several new metrics ~\cite{reddy2022dnsmos,kumar2023torchaudiosquim,NISQA} have been proposed which attempt to directly predict human quality assessment in a non-intrusive way, i.e.~where the clean speech reference is not required. These take the form of neural networks which are trained using vast datasets of distorted audio to predict a quality label assigned to the audio by the human assessors. \acp{SSSR} have been found to be useful feature representations for the prediction of audio quality \cite{close2023perceive}. \\
This paper comprises a system which builds on the authors' entry \cite{chime7entryclose} to the CHIME-7 challenge \ac{UDASE}~\cite{leglaive2023chime7} track. It attempts to address the problem of model adaption to real world data via a metric prediction \ac{GAN} based methodology. A non-intrusive \ac{GAN} discriminator is trained to predict multiple metrics including a MOS-related metric, as well as a traditional intrusive signal quality metric.  Historical training data from a conventional generator and an additional \emph{pseudo-generator} is used to augment the training data diversity. Then, during the training of the speech enhancement generator, inference of the multi-metric prediction discriminator is used to optimise the enhanced outputs towards the target metrics. In this way, metrics which are unable to be directly used as loss functions as well those which require access to a reference signal can be optimised.\\
The remainder of this paper is structured as follows. The  target metrics are described in  \autoref{metrics}. A description of the proposed Multi-CMGAN+/+ model is given in \autoref{sys_description}. Experimental setup and results are discussed in \autoref{exp_setup} and \autoref{results}, respectively. Finally, \autoref{conclusion} draws some conclusions from the findings of the paper.

\section{Speech Quality Metrics}\label{metrics}
Two speech quality metrics, \ac{PESQ} and \ac{DNSMOS}, are used as target metrics which the speech enhancement generator in our proposed system is trained to optimise towards.
\subsection{PESQ}
\acf{PESQ}~\cite{pesq} is a well-known intrusive speech quality measure. It takes a time domain signal of the clean reference audio $s[n]$ and the time-domain audio of the signal to be evaluated, e.g.~the noisy signal $x[n]$, and returns a value $Q_\mathrm{PESQ}$ between $1$ and $4.5$ which represents the quality of the test signal, higher meaning better quality:
\begin{equation}
    Q_\mathrm{PESQ} = \mathrm{PESQ}(s[n],x[n])
    \label{eq:pesq}
\end{equation}
The formulation of \ac{PESQ} is non-differentiable, so direct use of it as a loss function for training enhancement models is not possible.  
\subsection{DNSMOS}\label{quality_metrics}
\Acf{DNSMOS} \cite{reddy2022dnsmos} is a non-intrusive speech quality metric. It consists of a neural network which was trained to predict human \ac{MOS} ratings for speech signals. As it is non-intrusive, it is particularly useful for assessing the quality of real-world recordings such as in the CHiME-7 UDASE challenge testset, and was one of the evaluation metrics used in assessing the entries to the challenge.\\
For a input time domain speech signal $s[n]$ \ac{DNSMOS} estimates three values, being estimates of the well-known composite measure \cite{composite}:
\begin{equation}
    \left[Q_\mathrm{SIG},Q_\mathrm{BAK},Q_\mathrm{OVR}\right]= \mathrm{DNSMOS}(s[n]),
    \label{eq:dnsmos}
\end{equation}
where $Q_\mathrm{SIG}$, $Q_\mathrm{BAK}$ and $Q_\mathrm{OVR}$ are each values between $1$ and $5$ which represent the estimated speech quality, background noise quality and overall quality, respectively (higher values indicating better quality). In this work the non-neural implementation of \ac{DNSMOS} provided in the CHIME-7 baseline system is used.\\

\subsection{Non-intrusive Metric Prediction}
While \ac{DNSMOS} is a neural network meaning it is theoretically possible to backpropagate through it and use it directly in a loss function, it is not publicly available in this form. Similarly, the computation of \ac{PESQ} is non-differentiable, and requires access to a reference signal, meaning it cannot be used for most real-world scenarios.
In order to incorporate \ac{DNSMOS} and \ac{PESQ} in loss functions for speech enhancement in this work, a non-intrusive metric prediction discriminator~\cite{fu2021metricganu} is trained to create  differentiable `clones' of the metrics. This has the added benefit of allowing for an adversarial training of the metric prediction network in a \ac{GAN} setting~\cite{FuLTL19}.  In the following, $Q$ is used to represent one of these target metrics in \eqref{eq:pesq} and \eqref{eq:dnsmos} and $Q'$ is the respective value normalized between $0$ and $1$.

\section{Speech Enhancement System}\label{sys_description}
The overall architecture of the proposed system is based on the \ac{CMGAN} framework proposed in \cite{cmgan}, but with two extensions based on \cite{fu21_interspeech} and \cite{close2022}. The first extension is to train the discriminator $\mathcal{D}$ on a historical set of past generator outputs every epoch. The second extension is to train $\mathcal{D}$ to predict the metric score of noisy, clean and enhanced audio, as well as the output of a secondary pseudo-generator network $\mathcal{N}$ which is designed to increase the range of metric values observed by $\mathcal{D}$. This work introduces a new structure for $\mathcal{D}$ allowing it to predict multiple metrics at once, as well as a new input feature which is derived from a pre-trained \ac{SSSR} model.

\subsection{Conformer-based Speech Enhancement Generator}
    

    

\subsubsection{Conformer-based Generator Network Structure}
The conformer model generator $\mathcal{G}$ is based on the best performing CMGAN configuration in \cite{cmgan}. The network itself combines mapping and masking approaches for spectral speech enhancement, utilizing a conformer \cite{gulati2020conformer} based bottleneck.
The model's input are \ac{STFT} components of the complex-valued noisy audio, $\mathbf{X}_\mathrm{Re},\mathbf{X}_\mathrm{Im}$, with a reasonably high temporal resolution (hop size of $6$~ms with a $50$\% overlap, and a fast Fourier transform (FFT) length of $400$ samples). The output of the model are the enhanced real and imaginary \ac{STFT} components $\mathbf{\hat{S}}_\mathrm{Re}$ and $\mathbf{\hat{S}}_\mathrm{Im}$ from which the enhanced time domain audio $\hat{s}[n]$ is obtained by \ac{ISTFT}. Note that the time index $n$ is omitted for clarity in the following.

\subsubsection{Generator Loss Function}
The generator model $\mathcal{G}$ is trained with a multi-term loss function:
\begin{equation}
    \label{eq:g_loss}
    L_{\mathcal{G}} = {L_{\mathcal{G}_\mathrm{GAN}}} + {L_{\mathcal{G}_\mathrm{Time}}} + L_{\mathcal{G}_\mathrm{SI-SDR}}
\end{equation}
${L_{\mathcal{G}_\mathrm{GAN}}}$ minimises the distance 
\begin{equation}
\label{eq:gan_loss}
L_{\mathcal{G}_\mathrm{GAN}} = \mathbb{E}\left\{\Vert\mathcal{D}(\mathbf{\hat{S}}_\mathrm{FE}) - \mathbf{1}\Vert_2^2\right\},
\end{equation}
which represents an assessment of the enhanced signal by the metric Discriminator $\mathcal{D}$. $\mathcal{D}(\mathbf{\hat{S}}_\mathrm{FE})$ is the inference of the metric prediction discriminator $\mathcal{D}$, given the enhanced signal as input, which has an output of dimension $N_Q \times 1$ representing the $N_Q$ predicted normalised $Q'$ values of the target metrics, i.e.~$N_Q$ equals $3$ when using (\ref{eq:dnsmos}). The $\mathbf{1}$ vector in \eqref{eq:gan_loss}, also of length $N_Q$, represents the highest possible target metric values normalized between $0$ and $1$. Thus, the net effect of this loss term is to encourage $\mathcal{G}$ to maximise the predicted scores assigned to its outputs by $\mathcal{D}$. \\  ${L_{\mathcal{G}_\mathrm{Time}}}$ is a mean absolute error between the enhanced and clean time domain mixtures:
\begin{equation}
\label{eq:time_loss}
L_{\mathcal{G}_\mathrm{Time}} = \mathbb{E}\left\{||s - \hat{s}||_1\right\}.
\end{equation}
Finally, ${L_{\mathcal{G}_\mathrm{SISDR}}}$ is the \ac{SI-SDR} \cite{sisdr} loss
\begin{equation} 
\label{eq:DefSISDR}
 L_{\mathcal{G}_\mathrm{SISDR}} 
= - 10\log_{10} \frac{\left\Vert 
\frac{\langle \hat{{s}},{s}\rangle 
{s}}{\Vert {s}\Vert^{2}}\right\Vert^{2}}{\left\Vert\hat{{s}}-\frac{\langle \hat{{s}},{s}\rangle 
{s}}{\Vert {s}\Vert^{2}}\right\Vert^{2}}.\\
\end{equation}
With the exception of \eqref{eq:gan_loss}, all terms of $L_\mathcal{G}$ require access to clean label/reference audio $s$. 
\subsubsection{Block Processing for Longer Inputs} 
Due to the quadratic time-complexity of the transformer layers in the conformer models, processing long sequences can be unfeasible due to high memory requirements. Transformers are also typically unsuitable for continuous processing as the entire sequence is required to compute self-attention. 
To address these issues input signals are processed in overlapping blocks of $4$s for evaluation and inference as this has been shown to be in an optimal signal length for attention-based enhancement models \cite{tsllimits}. A $50$\% overlap with a Hann window is used to cross-fade each block with one an another. 
Models are trained with $4$s signal length limits \cite{tsllimits}.

\subsection{ Metric Estimation Discriminator}

The discriminator $\mathcal{D}$ part of the \ac{GAN} structure is trained to predict  three normalised speech quality metrics for a given input signal. Inference of $\mathcal{D}$ is used in \eqref{eq:gan_loss} as one of the loss terms of $\mathcal{G}$ and as the sole loss function of $\mathcal{N}$ in \eqref{eq:degen_loss}, enforcing an optimisation towards the target metrics.\\
We experiment with training $\mathcal{D}$ to predict each outputs of \ac{DNSMOS} (i.e $ Q_\mathrm{SIG},Q_\mathrm{BAK}$ or $Q_\mathrm{OVR}$), as well as \ac{PESQ} ($Q_\mathrm{PESQ}$).
\subsection{HuBERT Encoder Feature Representations}\label{hubert}
 Recent work in metric prediction \cite{sssr_quality1,close2023perceive}  shows that \acp{SSSR} are useful as feature extractors for capturing quality-related information about speech audio.
 As such, the proposed system makes use of the Hidden Unit BERT (HuBERT)~\cite{hubert} SSSR as a feature extractor for the metric prediction component of the proposed framework. HuBERT, like most \acp{SSSR} which take time domain signals as input, consists of two distinct network stages. The first stage, $\mathcal{H}_\mathrm{FE}(\cdot)$, comprises several 1D convolutional layers which map the input time-domain audio $s[n]$ into a $2$D representation $\mathbf{S}_{\mathrm{FE}}$. The second stage, $\mathcal{H}_\mathrm{OL}(\cdot)$, consists of a number of transformer 
 layers, which takes the output of the first stage $\mathbf{S}_{\mathrm{FE}}$ as input. The two representations $\mathbf{S}_\mathrm{FE}$ and $\mathbf{S}_\mathrm{OL}$ can thus be obtained from the HuBERT model:
 \begin{eqnarray}
    \label{eq:fe_transform}
    \mathbf{S}_\mathrm{FE} &=& \mathcal{H}_\mathrm{FE}({s}[n])\\
    \label{eq:ol_transform}
    \mathbf{S}_\mathrm{OL} &=& \mathcal{H}_\mathrm{OL}(\mathcal{H}_\mathrm{FE}({s}[n]))
\end{eqnarray}
Recent work in speech enhancement \cite{close2023perceive,close2023effect,10094705} have found that the outputs of HuBERT's encoder stage $\mathcal{H}_\mathrm{FE}(\cdot)$ are particularly useful for capturing quality-related information, outperforming the final transformer layer and weighted sums of each transformer output. The outputs of  $\mathcal{H}_\mathrm{FE}(\cdot)$ are 2D representations with dimensions $512 \times T$ where $T$ depends on the length of the input audio in seconds.  
The HuBERT model used in this work is trained on 960 hours of audio-book recordings from the LibriSpeech~\cite{7178964} dataset, sourced from the FairSeq GitHub repo\footnote{\url{https://github.com/facebookresearch/fairseq}}. This HuBERT encoder representation is used as a feature extractor, and its parameters are not updated during the training of the metric prediction network. 
\subsubsection{Discriminator Network Stucture}
 The discriminator network structure consists of $2$ \ac{BLSTM} layers followed by three parallel attention feed-forward layers with sigmoid activations, similar to the network proposed in \cite{sssr_quality1}. Each attention feed-forward layer outputs a single neuron which represents the prediction value of one of the three target metrics. 
The input to $\mathcal{D}$ is the output of the HuBERT feature encoder $\mathcal{H}_\mathrm{FE}(\cdot)$.
The output of $\mathcal{D}$ has dimension $B \times N_Q$ where $B$ is the batch size and each of $N_Q$ values represents a normalised predicted metric value.  
Note that inference of $\mathcal{D}$ is always non-intrusive, even when if one of it's target metrics such as \ac{PESQ} is intrusive. 

\subsubsection{Discriminator Loss Function}
Within each epoch, first the Discriminator $\mathcal{D}$ is trained on the current training elements:


\begin{flalign}
\label{eq:discrim_loss_MGplus}
\hspace*{-0.2cm}
L_{\mathcal{D},\mathrm{MG+}} 
= \mathbb{E}\{&(\mathcal{D}(\mathbf{S}_\mathrm{FE})-[Q'_1(s),... ,Q'_{N_Q}(s)])^2  \nonumber \\
&+ (\mathcal{D}(\mathbf{\hat{S}}_\mathrm{FE})-[Q'_1(\hat{s}),... ,Q'_{N_Q}(\hat{s})])^2 \nonumber \\
&+ (\mathcal{D}(\mathbf{X}_\mathrm{FE}) -
    [Q_1'(x),... ,Q'_{N_Q}(x)])^2 \nonumber \\
& +(\mathcal{D}(\mathbf{Y}_\mathrm{FE})
    -[Q_1'(y),... ,Q'_{N_Q}(y)])^2 \}
\end{flalign}
where $\mathbf{S}_\mathrm{FE}$, $\mathbf{X}_\mathrm{FE}$, $\mathbf{\hat{S}}_\mathrm{FE}$ and $\mathbf{Y}_\mathrm{FE}$ are HuBERT encoder representations, i.e.~after $\mathcal{H}_\mathrm{FE}(\cdot)$, of the clean signal $s$, the noisy signal $x$, the signal enhanced by $\mathcal{G}$, $\hat{s}$, and the signal as enhanced by $\mathcal{N}$, $y$. $Q'_1(\cdot),Q'_2(\cdot)$ and $Q'_3(\cdot)$ are the true target metric scores of the input audio, normalized between $0$ and $1$. Please note that the $Q'$ vectors in (\ref{eq:discrim_loss_MGplus}) can be shorter than $3$ if less than $N_Q=3$ metrics are considered.
This is followed by a historical training stage, where $\mathcal{D}$ is trained to predict the metric scores from past outputs of the generative networks $\mathcal{G}$ and $\mathcal{N}$. 
\subsubsection{Historical Training}
The training procedure of $\mathcal{D}$ uses historical training data as first proposed in the MetricGAN+ framework~\cite{fu21_interspeech}. In this stage, a sample of enhanced audio output from past epochs of $\mathcal{G}$ and $\mathcal{N}$ are used to train $\mathcal{D}$. This aim of this is to widen prevent $\mathcal{D}$ from `forgetting' how to assess audio which is dissimilar to the current outputs of the enhancement network. In each epoch, $\mathcal{D}$ is trained using a randomly selected $10\%$ of the outputs of the generator models from past epochs. 

\subsection{Metric Data Augmentation Pseudo-Generator}
As first proposed in \cite{close2022}, a secondary speech enhancement network $\mathcal{N}$ is trained, and its outputs $y$ used to train the metric prediction discriminator $\mathcal{D}$ (last term in \eqref{eq:discrim_loss_MGplus}) . This model is trained solely using the GAN loss in (\ref{eq:gan_loss}), similar to the original MetricGAN framework:
\begin{equation}
\label{eq:degen_loss}
L_{\mathcal{N}_\mathrm{GAN}} = \mathbb{E}\{\Vert\mathcal{D}(\mathbf{Y}_\mathrm{FE}) - \mathbf{1} w\Vert_2^2\}
\end{equation}
where $w$ is a hyperparameter value which corresponds to the target normalised DNSMOS score for which the output audio of $\mathcal{N}$ is being trained to obtain. Following on from prior work~\cite{chime7entryclose}, here we fix the value of $w$ at $1$ meaning that $\mathcal{N}$ is trained to enhance relative to the target metrics, rather than to 'de-enhance' with a lower value of $w$.\\
$\mathcal{N}$s network structure is based on the original MetricGAN enhancement model, consisting of a \ac{BLSTM} which operates on a magnitude spectrogram representation of the input, followed by 3 linear layers. Its output is a magnitude mask which is multiplied by the input noisy spectrogram to produce an enhanced spectrogram $\mathbf{Y}_\mathrm{SPEC}$. A time domain signal ${y}[n]$ is constructed by the overlap-add method using the original noisy phase. \\
\vspace{-1cm}
\section{Experiments}\label{exp_setup}
\subsection{Training Setup}
The framework is trained on simulated labelled data from the LibriMix~\cite{cosentino2020librimix} for $200$ epochs, following a similar dataloading system as in \cite{leglaive2023chime7} generating mixtures of a single speaker with noise.
The labelled LibriMix training set consists of $33900$ clean/noisy audio pairs, with the clean speech sourced from the LibriSpeech~\cite{7178964} dataset and the added noise from WHAM!~\cite{wichern2019wham} dataset.
\\
Each epoch, $300$  samples from the training set are randomly selected. These are first used to train the metric prediction Discriminator $\mathcal{D}$ using \eqref{eq:discrim_loss_MGplus}. This is followed by the training of $\mathcal{D}$ on the historical set. Then the $300$ random samples are used to  train $\mathcal{N}$  using inference of $\mathcal{D}$ with \eqref{eq:degen_loss}, followed finally by the training of $\mathcal{G}$ using \eqref{eq:g_loss} which also uses inference of $\mathcal{D}$.\\
Different combinations of the DNSMOS terms and PESQ are experimented with as the three target metrics for $\mathcal{D}$  by setting 
 each of $Q_1,Q_2,Q_3$ in \eqref{eq:discrim_loss_MGplus} to be $Q_\mathrm{PESQ}, Q_\mathrm{SIG},Q_\mathrm{BAK}$ or $Q_\mathrm{OVR}$.\\
The proposed models are evaluated on the CHiME7 UDASE task \cite{leglaive2023chime7} evaluation sets. These are a real world unlabelled set consisting of CHIME5 recordings which are evaluated using DNSMOS and a simulated labelled set consisting of reverberant LibriMix audio which are evaluated using \ac{SI-SDR}. The proposed system is compared to our prior entry to the CHiME7 \ac{UDASE} challenge \cite{chime7entryclose}, as well as the challenge baselines \cite{leglaive2023chime7}. Source code will be available at \footnote{\url{https://github.com/leto19/MultiMetricGANplusplus}}. 
\section{Results}\label{results}
\autoref{tab:chime} shows the results of the proposed framework in terms of DNSMOS on the CHiME-7 \ac{UDASE} task real evaluation set.
\begin{table}[!ht]
\caption{DNSMOS results on CHiME5 eval set.\label{tab:chime}}
\centering
\resizebox{\columnwidth}{!}{%
\begin{tabular}{l|c|ccc}
\hline
\textbf{Model}     & \textbf{$Q_1,Q_2,Q_3$} & \multicolumn{1}{l}{\textbf{OVR}} & \multicolumn{1}{l}{\textbf{BAK}} & \multicolumn{1}{l}{\textbf{SIG}} \\
\hline
\textit{unprocessed}          &          --              & \textit{2.84}                    & \textit{2.92}                    & \textit{3.48}                    \\
Sudo rm -rf~\cite{Tzinis_2020}                  & --                        & 2.88                             & 3.59                             & 3.33                             \\
RemixIT~\cite{Tzinis_2022} w/VAD              &       --                 & 2.84                             & 3.62                             & 3.28                             \\
CMGAN+/+~\cite{chime7entryclose}        & SIG                    & 3.29                             & 3.85                    & \textbf{3.76}                             \\
\hline
Multi-CMGAN+/+         & SIG/BAK/OVR      &\textbf{3.42} &3.86 &3.56      \\
Multi-CMGAN+/+          & SIG/BAK/PESQ      &3.08&3.78 &3.41            \\
Multi-CMGAN+/+         & SIG/OVR/PESQ      &2.80 &3.62 &3.19         \\
Multi-CMGAN+/+         & BAK/OVR/PESQ &3.12 &\textbf{3.86} &3.49            \\
\end{tabular}
}
\end{table}
The proposed systems significantly outperform the baseline systems in all measures, while also outperforming the author's prior work CMGAN+/+ in terms of OVR and BAK. However, CMGAN+/+ still outperforms the proposed system in terms of SIG, which is the only metric it is optimized towards.

\begin{table}[!h]
\caption{SI-SDR results on the reverberant LibriCHiME eval set.}
\label{tab:reverb-librichime}
\centering
\resizebox{\columnwidth}{!}{%
\begin{tabular}{l|c|c}
\hline
\textbf{Model}             & $Q_1,Q_2,Q_3$                          & \textbf{SI-SDR (dB)} \\
\hline
\textit{unprocessed}             &-- &              \textit{6.59}                             \\
Sudo rm -rf~\cite{Tzinis_2020}  &-- & 7.8                                       \\
RemixIT~\cite{Tzinis_2022} w/ VAD          &-- & \textbf{10.05}                            \\ \hline
CMGAN+/+        & SIG             &           4.71       \\
\hline
Multi-CMGAN+/+         & SIG/BAK/OVR      &     3.36      \\
Multi-CMGAN+/+         & SIG/BAK/PESQ      &    4.47      \\
Multi-CMGAN+/+         & SIG/OVR/PESQ      &    0.09        \\ 
Multi-CMGAN+/+         & BAK/OVR/PESQ      &6.95\\
\end{tabular}
}
\end{table}
\autoref{tab:reverb-librichime} shows the results of the proposed framework in terms of \ac{SI-SDR} on the  CHiME-7 \ac{UDASE} task simulated evaluation set. Here, the weaknesses of the proposed system relative to the CHiME-7 baseline systems is apparent, with our proposed framework significantly degrading the input with the exception of the model which does \emph{not} optimise the SIG component of DNSMOS. 
\vspace{-0.5cm}
\section{Conclusion}\label{conclusion}
In this work a \ac{GAN} framework utilising a multi-metric prediction discriminator is introduced. A number of combinations of target metric for this prediction network are experimented with, and improved performance on test set consisting of real data is shown. However a degradation in performance on a simulated testset is also shown, suggesting a significant distortion in the enhanced outputs of the proposed system.  



\let\OLDthebibliography\thebibliography
\renewcommand\thebibliography[1]{
  \OLDthebibliography{#1}
  \setlength{\itemsep}{0pt plus 0.7ex}
}

\bibliographystyle{IEEEtran}
\bibliography{strings,refs}

\end{document}